\documentclass[copyright,creativecommons]{eptcs}
\providecommand{\event}{ACL2 Workshop 2025}
\usepackage{underscore, latexsym,pifont,amsmath,amstext,amssymb,amsfonts,makeidx,url}
\title{A Formalization of Elementary Linear Algebra: Part II}
\author{David M.~Russinoff
\email{david@russinoff.com}
}
\def\titlerunning{A Formalization of Linear Algebra: Part II}
\def\authorrunning{D.M. Russinoff}

\begin{document}
\maketitle

\begin{abstract}
This is the second installment of an exposition of an ACL2 formalization of elementary linear algebra.  It extends the results
of Part I, which covers the algebra of matrices over a commutative ring, but focuses on
aspects of the theory that apply only to matrices over a field: elementary row reduction and its
application to the computation of matrix inverses and the solution of simultaneous systems of linear equations.
\end{abstract}

\section{Introduction}

This is the second installment of an exposition of an ACL2 formalization of elementary linear algebra.  Part I \cite{linear1}, which is
also included in this workshop, covers the algebra of matrices over a commutative ring with unity and their determinants.
In this sequel, we focus on aspects of the theory of matrices that apply only to matrices over a field, i.e., depend on
the existence of a multiplicative inverse operator.  These include row reduction and its application to
matrix invertibility and the solution of systems of linear equations.  In an anticipated Part~III, all of these results will be applied
to the study of abstract vector spaces and linear transformations.

The proof scripts supporting both papers reside in the same directory, {\tt books/projects/linear/}.  As described in \cite{linear1},
the abstract definition of a ring is formalized in the file {\tt ring.lisp} by a set of constrained encapsulated functions:
a predicate {\tt rp} that recognizes ring elements, the binary addition and multiplication operations {\tt r+} and {\tt r*},
the corresponding identity constants {\tt r0} and {\tt r1}, and the additive inverse operator {\tt r-}.  The notion of a field
is similarly defined by an encapsulation in the file {\tt field.lisp}, in which these functions are renamed {\tt fp}, {\tt f+},
{\tt f*}, etc.:

\begin{small}
\begin{verbatim}
  (encapsulate (((fp *) => *)                   ;field element recognizer
                ((f+ * *) => *) ((f* * *) => *) ;addition and multiplication
                ((f0) => *) ((f1) => *)         ;identities
                ((f- *) => *) ((f/ *) => *))    ;inverses 
    (local (defun fp (x) (rationalp x)))
    (local (defun f+ (x y) (+ x y)))
    (local (defun f* (x y) (* x y)))
    (local (defun f0 () 0))
    (local (defun f1 () 1))
    (local (defun f- (x) (- x)))
    (local (defun f/ (x) (/ x)))
    ;; Closure:

    ...
  
    ;; Multiplicative inverse:
    (defthm fpf/
      (implies (and (fp x) (not (equal x (f0)))) (fp (f/ x))))
    (defthm f*inv
      (implies (and (fp x) (not (equal x (f0)))) (equal (f* x (f/ x)) (f1)))))
\end{verbatim}
\end{small}
The only other difference between the two encapsulations is the inclusion here of the multiplicative inverse {\tt f/}
along with two constraining axioms, appended to the constraints adapted from the ring axioms.

Informally we shall refer to the field {\tt F} that is characterized by this encapsulation.  When our intention is clear,
the identity elements {\tt (f0)} and {\tt (f1)} will be abbreviated as 0 and 1.
Clearly, all properties of rings hold for fields as well.  Thus, all definitions and theorems that appear in
{\tt ring.lisp}, {\tt rmat.lisp}, and {\tt rdet.lisp} have analogs in the corresponding files {\tt field.lisp},
{\tt fmat.lisp}, and {\tt fdet.lisp}.  In particular, the predicate {\tt flistnp} recognizes a list of specified length of
elements of {\tt F}, called an {\it flist}; {\tt fmatp} recognizes a matrix over {\tt F};
and {\tt fdet} computes its determinant.  In principle, all results in the latter set of files could be derived
by functional instantiation from the corresponding events in the former, but we found it more expedient to reproduce the
proofs, simply by selectively replacing occurrences of the character {\tt r} with {\tt f}.  An additional file,
{\tt reduction.lisp}, contains the results reported in this paper.  In describing these results, we assume the reader
is familiar with Part~I and is aware of the renaming convention.

In Section~\ref{reduction}, we define the notion of a reduced row-echelon matrix and develop a procedure that converts an arbitrary matrix
to reduced row-echelon form.  An equivalent procedure, based on matrix multiplication, is also defined.  This leads to a
criterion for invertibility of a square matrix and a method for computing inverses.  A second method of matrix inversion, based 
on determinants and the classical adjoint, is derived from the results of Part~I.

Section~\ref{equations} addresses the solution of systems of linear equations, mainly as an application of row reduction.
We derive algorithmic tests for solvability and uniqueness of the solution, as well as a formula that computes the
solution in the uniquely solvable case. For the special case of an invertible square coefficient matrix, we prove
Cramer's Rule, an alternative formula based on determinants.  In the general solvable case, we show that the
solution set is infinite and establish a test that identifies solutions.  In Part~III, this will lead to a formula that generates
a basis for the solution space of a homogeneous system of equations.

All of these results, which are stated and proved in the context of an abstract field, may be applied to any concrete field of 
interest through functional instantiation.  Eventually, we plan to apply the theory to algebraic number fields.  Of course, in
their abstract formulation based on constrained functions, the definitions are not executable.  For the immediate purposes of
illustration and testing, however, all functions defined in {\tt field.lisp}, {\tt fmat.lisp}, {\tt fdet.lisp}, and {\tt reduction.lisp}
have been adapted to the field of rational numbers as executable functions, which are listed in the file {\tt rational.lisp}.

\section{Row Reduction}\label{reduction}

\subsection{Reduced Row-Echelon Form}

A {\it reduced row-echelon} matrix may be characterized as follows:

\begin{itemize}
\item [(1)] Every all-zero row is preceded by every nonzero row;
\item [(2)] The first nonzero entry of each nonzero row is 1, and every other entry in the same
column is 0;
\item [(3)] The column of the leading 1 in the {\tt i}th nonzero row is an increasing function of {\tt i}.
\end{itemize}
The formalization of this definition requires several auxiliary functions.  First, we define the index of the leading nonzero
entry of a nonzero row {\tt r}:

\begin{small}
\begin{verbatim}
(defun first-nonzero (r)
  (if (consp r)
      (if (= (car r) (f0))
          (1+ (first-nonzero (cdr r)))
        0)        
    ()))
\end{verbatim}
\end{small}
In the following, we assume that {\tt a} is an {\tt m}$\times${\tt n} matrix.  Starting with row {\tt k}, where 0
$\leq$ {\tt k} $\leq$ {\tt m}, find the row of {\tt a} with nonzero entry of least index, or return {\tt NIL} if all rows beyond
the first {\tt k} are 0:

\begin{small}
\begin{verbatim}
(defun row-with-nonzero-at-least-index (a m k)
  (if (and (natp k) (natp m) (< k m))
      (let ((i (row-with-nonzero-at-least-index a (1- m) k)))
        (if (or (flist0p (nth (1- m) a))
                (and i (<= (first-nonzero (nth i a)) (first-nonzero (nth (1- m) a)))))
            i
          (1- m)))
    ()))
\end{verbatim}
\end{small}
Given {\tt j} $<$ {\tt n} and {\tt k} $<$ {\tt m}, check that {\tt (entry k j a)} = 1 and that all other entries in column {\tt k}
are 0:

\begin{small}
\begin{verbatim}
(defun column-clear-p (a k j m)
  (if (zp m) t
    (and (= (nth j (nth (1- m) a)) (if (= (1- m) k) (f1) (f0)))
         (column-clear-p a k j (1- m)))))
\end{verbatim}
\end{small}
Given {\tt k} $\leq$ {\tt m}, check that the first {\tt k} rows of {\tt a} form a reduced row-echelon matrix:

\begin{small}
\begin{verbatim}
  (defun row-echelon-p-aux (a m k)
    (if (zp k) t
      (and (row-echelon-p-aux a m (1- k))
           (let ((i (row-with-nonzero-at-least-index a m (1- k))))
             (or (null i)
                 (and (= i (1- k))
                      (column-clear-p a i (first-nonzero (nth i a)) m)))))))
\end{verbatim}
\end{small}
Finally, check that {\tt a} is a reduced row-echelon matrix:

\begin{small}
\begin{verbatim}
  (defund row-echelon-p (a) (row-echelon-p-aux a (len a) (len a)))
\end{verbatim}
\end{small}

\subsection{Conversion to Reduced Row-Echelon Form}\label{conversion}

We shall develop a procedure that converts an arbitrary {\tt m}$\times${\tt n} matrix {\tt a} to reduced row-echelon form by a sequence of
{\it elementary row operations} of three types:\medskip

(1) Multiply row {\tt k} by a scalar {\tt c}:

\begin{small}
\begin{verbatim}
       (defund ero1 (a c k) (replace-row a k (flist-scalar-mul c (nth k a))))
\end{verbatim}
\end{small}

(2) Add a scalar multiple of row {\tt j} to row {\tt k}, where {\tt j} $\neq$ {\tt k}:

\begin{small}
\begin{verbatim}
       (defund ero2 (a c j k)
         (replace-row a k (flist-add (flist-scalar-mul c (nth j a)) (nth k a))))
\end{verbatim}
\end{small}

(3) Interchange rows {\tt j} and {\tt k}, where {\tt j} $\neq$ {\tt k}:

\begin{small}
\begin{verbatim}
       (defund ero3 (a j k) (replace-row (replace-row a k (nth j a)) j (nth k a)))
\end{verbatim}
\end{small}
Under the assumption that {\tt (entry k j a)} = 1, the following function applies {\tt ero2} to clear all other entries in column {\tt j}
by adding the appropriate multiple of row {\tt k} to each of the other rows:

\begin{small}
\begin{verbatim}
  (defun clear-column (a k j m)
    (if (zp m) a
      (if (= (1- m) k)
          (clear-column a k j (1- m))
        (clear-column (ero2 a (f- (nth j (nth (1- m) a))) k (1- m))
                      k j (1- m)))))
\end{verbatim}
\end{small}
Assume the first {\tt k} rows of {\tt a} are in reduced row-echelon form, i.e., {\tt (row-echelon-p-aux a m k)} = {\tt T}, where
{\tt k} $<$ {\tt m}, and that {\tt i} = {\tt (row-with-nonzero-at-least-index a m k)} $\neq$ {\tt NIL}.
Let {\tt j} = {\tt (first-nonzero (nth i a))}.  The following function performs the next step of the reduction, producing
a matrix {\tt a'} satisfying {\tt (row-echelon-p-aux a' m (1+ k))}:

\begin{small}
\begin{verbatim}
  (defund row-reduce-step (a m k i j)
    (clear-column (ero3 (ero1 a (f/ (nth j (nth i a))) i)
                        i k)
                  k j m))
\end{verbatim}
\end{small}
The function {\tt row-reduce} converts {\tt a} to a reduced row-echelon matrix, using an auxiliary function that completes the
conversion under the assumption {\tt (row-echelon-p-aux a m k)}, where 0 $\leq$ {\tt k} $\leq$ {\tt m}:

\begin{small}
\begin{verbatim}
  (defun row-reduce-aux (a m k)
    (let ((i (row-with-nonzero-at-least-index a m k)))
      (if (and (natp k) (natp m) (< k m) i)
          (row-reduce-aux (row-reduce-step a m k i (first-nonzero (nth i a)))
                          m (1+ k))
        a)))

  (defund row-reduce (a) (row-reduce-aux a (len a) 0))
\end{verbatim}
\end{small}
The following confirms that this procedure produces the desired result:

\begin{small}
\begin{verbatim}
  (defthmd row-echelon-p-row-reduce
    (implies (and (natp m) (natp n) (fmatp a m n))
             (row-echelon-p (row-reduce a))))
\end{verbatim}
\end{small}
We also note that row reduction does not alter a reduced row-echelon matrix:

\begin{small}
\begin{verbatim}
  (defthmd row-reduce-row-echelon-p
    (implies (and (posp m) (posp n) (fmatp a m n) (row-echelon-p a))
             (equal (row-reduce a) a)))
\end{verbatim}
\end{small}

As an example, consider the following $4\times 5$ matrix {\tt (a0)}:

\begin{small}
\begin{verbatim}
  DM !>(defun a0 () '((0 -3 -6 4 9) (-1 -2 -1 3 1) (-2 -3 0 3 -1) (1 4 5 -9 -7)))
\end{verbatim}
\end{small}
In the first step in the row reduction of {\tt (a0)}, row 1 is divided by its leading nonzero entry, -1, and interchanged with row 0.  The other entries in column 0 are then cleared:

\begin{small}
\begin{verbatim}
  DM !>(row-reduce-step (a0) 4 0 1 0)
  ((1  2  1 -3 -1)
   (0 -3 -6  4  9)
   (0  1  2 -3 -3)
   (0  2 4  -6 -6))
\end{verbatim}
\end{small}
Row reduction of {\tt (a0)} requires three executions of {\tt row-reduce-step}:

\begin{small}
\begin{verbatim}
  DM !>(row-reduce (a0))
  ((1  0 -3  0  5)
   (0  1  2  0 -3)
   (0  0  0  1  0)
   (0  0  0  0  0))
\end{verbatim}
\end{small}

We define the {\it row rank} of {\tt a} to be the number of nonzero rows of {\tt (row-reduce a)}:

\begin{small}
\begin{verbatim}
  (defun num-nonzero-rows (a)
    (if (consp a)
        (if (flist0p (car a)) 0 (1+ (num-nonzero-rows (cdr a))))
      0))
      
  (defun row-rank (a) (num-nonzero-rows (row-reduce a)))
\end{verbatim}
\end{small}
Note that {\tt (row-reduce (a0))} has 3 nonzero rows:

\begin{small}
\begin{verbatim}
  DM !>(row-rank (a0))
  3
\end{verbatim}
\end{small}
Obviously, the row rank of an {\tt m}$\times${\tt n} matrix cannot exceed {\tt m}:

\begin{small}
\begin{verbatim}
  (defthmd row-rank<=m
    (implies (and (fmatp a m n) (posp m) (posp n))
             (<= (row-rank a) m)))
\end{verbatim}
\end{small}
Nor can the row rank exceed {\tt n}.  To see this, consider the list of indices of the leading 1s of the nonzero rows of a reduced row-echelon matrix {\tt a}:

\begin{small}
\begin{verbatim}
(defun lead-inds (a)
  (if (and (consp a) (not (flist0p (car a))))
      (cons (first-nonzero (car a)) (lead-inds (cdr a)))
    ()))

  DM !>(lead-inds (row-reduce (a0)))
  (0 1 3)
\end{verbatim}
\end{small}
Clearly, the length of {\tt (lead-inds a)} is the number of nonzero rows of {\tt a}.
Furthermore, {\tt (lead-inds a)} is a strictly increasing sublist of {\tt (ninit t)}.
It follows that {\tt (len (lead-inds a))} $\leq$ {\tt n}.  Consequently, the row rank of {\tt a} is
bounded by {\tt n}:

\begin{small}
\begin{verbatim}
  (defthmd row-rank<=n
    (implies (and (fmatp a m n) (posp m) (posp n))
             (<= (row-rank a) n)))
\end{verbatim}
\end{small}
We also note that if {\tt (row-rank a)} = {\tt n}, then {\tt (lead-inds a)} is an increasing sublist of {\tt (ninit n)}
of length {\tt n}, which implies that the two lists are equal:

\begin{small}
\begin{verbatim}
  (defthmd lead-inds-ninit
    (implies (and (fmatp a m n) (posp m) (posp n)
                  (row-echelon-p a) (= (row-rank a) n))
             (equal (lead-inds a) (ninit n))))
\end{verbatim}
\end{small}

Along with row reduction, there is an obvious analogous notion of {\it column reduction} and a corresponding definiton of the
{\it column rank} of a matrix, which may alternatively be defined as the row rank of its transpose.  As we shall show in Part~III in the
context of vector spaces, the row and column ranks of a matrix are always equal.

\subsection{Row Reduction as Matrix Multiplication}

Once we have identified the sequence of operations required to derive the reduced row-echelon form of an {\tt m} $\times$ {\tt n} matrix {\tt a},
an alternative derivation may be achieved by applying the same operations to the {\tt m} $\times$ {\tt m} identity matrix and right-multiplying 
the result by {\tt a}.  To this end, a row operation is encoded as a list of length 3 or 4; the first member indicates the operation type (1, 2, or 3 as listed in the preceding subsection), and
the others are the parameters of the operation.  The following predicate characterizes an encoding of a row operation on a matrix of {\tt m} rows:

\begin{small}
\begin{verbatim}
(defund row-op-p (op m)
  (and (true-listp op)
       (case (car op)
         (1 (and (= (len op) 3) (fp (cadr op)) (not (= (cadr op) (f0)))
                 (natp (caddr op)) (< (caddr op) m)))
         (2 (and (= (len op) 4) (fp (cadr op))
                 (natp (caddr op)) (< (caddr op) m)
                 (natp (cadddr op)) (< (cadddr op) m)
                 (not (= (caddr op) (cadddr op)))))
         (3 (and (= (len op) 3) (natp (cadr op)) (< (cadr op) m)
                 (natp (caddr op)) (< (caddr op) m))))))         
\end{verbatim}
\end{small}
The function {\tt apply-row-op} applies an encoded row operation to a matrix:

\begin{small}
\begin{verbatim}
(defund apply-row-op (op a)
  (case (car op)
    ;(apply-row-op (list 1 c k) a) = (ero1 a c k)
    (1 (ero1 a (cadr op) (caddr op)))             
    ;(apply-row-op (list 2 c j k) a) = (ero2 a c j k)
    (2 (ero2 a (cadr op) (caddr op) (cadddr op))) 
    ;(apply-row-op (list 3 j k) a) = (ero3 a j k)
    (3 (ero3 a (cadr op) (caddr op)))))           
\end{verbatim}
\end{small}
A list of row operations is identified in the obvious way:

\begin{small}
\begin{verbatim}
(defun row-ops-p (ops m)
  (if (consp ops)
      (and (row-op-p (car ops) m)
           (row-ops-p (cdr ops) m))
    (null ops)))
\end{verbatim}
\end{small}
The function {\tt apply-row-ops} applies a list of operations in sequence from left to right:

\begin{small}
\begin{verbatim}
(defun apply-row-ops (ops a)
    (if (consp ops)
        (apply-row-ops (cdr ops) (apply-row-op (car ops) a))
      a))
\end{verbatim}
\end{small}
By examining the definitions of {\tt row-reduce} and its auxiliary functions, we construct the list of encodings of the operations that reduce a
matrix {\tt a} to reduced row-echelon form.  The next four functions encode the lists of operations performed by {\tt clear-column},
{\tt row-reduce-step}, {\tt row-reduce-aux}, and {\tt row-reduce}, respectively:

\begin{small}
\begin{verbatim}
  (defun clear-column-ops (a k j m)
    (if (zp m) ()
      (if (= k (1- m))
          (clear-column-ops a k j (1- m))
        (cons (list 2 (f- (nth j (nth (1- m) a))) k (1- m))
              (clear-column-ops (ero2 a (f- (nth j (nth (1- m) a))) k (1- m))
                                k j (1- m))))))
                                
  (defund row-reduce-step-ops (a m k i j)
    (cons (list 1 (f/ (nth j (nth i a))) i)
          (cons (list 3 i k)
                (clear-column-ops (ero3 (ero1 a (f/ (nth j (nth i a))) i) i k)
                                  k j m))))
                                  
  (defun row-reduce-aux-ops (a m k)
    (let* ((i (row-with-nonzero-at-least-index a m k))
           (j (and i (first-nonzero (nth i a)))))
      (if (and (natp k) (natp m) (< k m) i)
          (append (row-reduce-step-ops a m k i j)
                  (row-reduce-aux-ops (row-reduce-step a m k i j) m (1+ k)))
                  
  (defund row-reduce-ops (a) (row-reduce-aux-ops a (len a) 0))
\end{verbatim}
\end{small}
The correctness of this encoding procedure is confirmed by the following:

\begin{small}
\begin{verbatim}
  (defthmd apply-row-reduce-ops
    (implies (and (fmatp a m n) (posp m) (posp n))
             (equal (apply-row-ops (row-reduce-ops a) a)
                    (row-reduce a))))
\end{verbatim}
\end{small}

Returning to the example of Subsection~\ref{conversion}, we find that the first step in the row reduction of {\tt (a0)} involves five elementary operations:

\begin{small}
\begin{verbatim}
  DM !>(row-reduce-step-ops (a0) 4 0 1 0)
  ((1 -1 1) (3 1 0) (2 -1 0 3) (2 2 0 2) (2 0 0 1))

  DM !>(apply-row-ops '((1 -1 1) (3 1 0) (2 -1 0 3) (2 2 0 2) (2 0 0 1)) (a0))
  ((1  2  1 -3 -1)
   (0 -3 -6  4  9)
   (0  1  2 -3 -3)
   (0  2  4 -6 -6))
\end{verbatim}
\end{small}
The reader may wish to compute  {\tt (row-reduce-ops (a0))}, a list of length 15, and check that the lemma  {\tt apply-row-reduce-ops} holds in this case.

The {\tt m}$\times${\tt m} {\it elementary matrix} corresponding to a row operation is defined to be the result of applying the
operation to the {\tt m}$\times${\tt m} identity matrix:

\begin{small}
\begin{verbatim}
  (defund elem-mat (op m) (apply-row-op op (id-fmat m)))
\end{verbatim}
\end{small}
Application of a row operation is equivalent to left multiplication by the corresponding elementary matrix:

\begin{small}
\begin{verbatim}
  (defthmd elem-mat-row-op
    (implies (and (fmatp a m n) (row-op-p op m) (posp m) (posp n))
             (equal (fmat* (elem-mat op m) a) (apply-row-op op a))))
\end{verbatim}
\end{small}
The product of the list of elementary matrices associated with the row reduction of a matrix is computed recursively by the
function {\tt row-reduce-mat}:

\begin{small}
\begin{verbatim}
  (defund row-ops-mat (ops m)
    (if (consp ops)
        (fmat* (row-ops-mat (cdr ops) m) (elem-mat (car ops) m))             
      (id-fmat m)))
      
  (defund row-reduce-mat (a) (row-ops-mat (row-reduce-ops a) (len a)))
\end{verbatim}
\end{small}
It follows from {\tt elem-mat-row-op} by induction that applying a sequence {\tt ops} of row operations to {\tt a} is equivalent to multiplication of {\tt a} by
{\tt (row-ops-mat ops m)}:

\begin{small}
\begin{verbatim}
(defthmd fmat*-row-ops-mat
  (implies (and (fmatp a m n) (posp m) (posp n)
                (row-ops-p ops m))
           (equal (fmat* (row-ops-mat ops m) a)
                  (apply-row-ops ops a))))
\end{verbatim}
\end{small}
In particular, by {\tt apply-row-reduce-ops}, row reduction of {\tt a} is equivalent to multiplication by {\tt (row\-reduce-mat a)}:

\begin{small}
\begin{verbatim}
  (defthmd row-ops-mat-row-reduce
    (implies (and (fmatp a m n) (posp m) (posp n))
             (equal (fmat* (row-reduce-mat a) a) (row-reduce a))))
\end{verbatim}
\end{small}
In our example, the product of the 15 elementary matrices corresponding to {\tt (row-reduce-ops (a0))} is

\begin{small}
\begin{verbatim}
  DM !>(row-reduce-mat (a0))
  (( 3/5 -3/5 -1/5 0)
   (-3/5  8/5 -4/5 0)
   (-1/5  6/5 -3/5 0)
   (  0    5   -2  1))
\end{verbatim}
\end{small}
The conclusion of the lemma {\tt row-ops-mat-row-reduce} may be readily verified for this case.

\subsection{Invertibility}\label{invert}

In this subsection, we focus on square matrices.  Given an {\tt n}$\times${\tt n} matrix {\tt a}, we seek an {\it inverse} of {\tt a}, i.e., an {\tt n}$\times${\tt n} matrix
{\tt b} such that 

\begin{small}
\begin{verbatim}
  (fmat* a b) = (fmat* b a) = (id-fmat n).
\end{verbatim}
\end{small}
If such a matrix exists, then it is unique in the strong sense that it is the only left or right inverse of {\tt a}.  For example, if
{\tt (fmat* c a)} = {\tt (id-fmat n)}, then

\begin{small}
\begin{verbatim}
  c = (fmat* c (id-fmat n))
    = (fmat* c (fmat* a b))
    = (fmat* (fmat* c a) b))
    = (fmat* (id-fmat n) b))
    = b,
\end{verbatim}
\end{small}
and the same conclusion similarly follows from the assumption {\tt (fmat* a c)} = {\tt (id-fmat n)}.  Thus, we have

\begin{small}
\begin{verbatim}
  (defthm inverse-unique
    (implies (and (fmatp a n n) (fmatp b n n) (fmatp c n n) (posp n)
                  (= (fmat* a b) (id-fmat n)) (= (fmat* b a) (id-fmat n))
                  (or (= (fmat* a c) (id-fmat n)) (= (fmat* c a) (id-fmat n))))
             (equal c b)))
\end{verbatim}
\end{small}
Every elementary matix has an inverse:

\begin{small}
\begin{verbatim}
  (defund invert-row-op (op)
    (case (car op)
      (1 (list 1 (f/ (cadr op)) (caddr op)))
      (2 (list 2 (f- (cadr op)) (caddr op) (cadddr op)))
      (3 op)))
      
  (defthmd fmat*-elem-invert-row-op
    (implies (and (row-op-p op n) (posp n))
             (and (equal (fmat* (elem-mat (invert-row-op op) n) (elem-mat op n))
                         (id-fmat n))
                  (equal (fmat* (elem-mat op n) (elem-mat (invert-row-op op) n))                              
                         (id-fmat n)))))
\end{verbatim}
\end{small}
Consequently, every product of elementary matrices has an inverse:

\begin{small}
\begin{verbatim}
  (defun invert-row-ops (ops)
    (if (consp ops)
        (append (invert-row-ops (cdr ops)) (list (invert-row-op (car ops))))
      ()))
      
  (defthmd invert-row-ops-mat
    (implies (and (row-ops-p ops n) (posp n))
             (and (equal (fmat* (row-ops-mat (invert-row-ops ops) n)
                                (row-ops-mat ops n))
                         (id-fmat n))
                  (equal (fmat* (row-ops-mat ops n)
                                (row-ops-mat (invert-row-ops ops) n))               
                         (id-fmat n)))))
\end{verbatim}
\end{small}

We shall show that {\tt a} has an inverse iff {\tt (row-rank a) = n} and that in this case, the inverse of {\tt a} is
{\tt (row-reduce-mat a)}.  Thus, we define

\begin{small}
\begin{verbatim}
  (defund invertiblep (a n) (= (row-rank a) n))
\end{verbatim}
\end{small}
and  
  
\begin{small}
\begin{verbatim}
  (defund inverse-mat (a) (row-reduce-mat a))
\end{verbatim}
\end{small}
First we note that as a consequence of {\tt  lead-inds-ninit}, if {\tt (invertiblep a n)}, then {\tt (row-reduce a)} = {\tt (id-fmat n)}:

\begin{small}
\begin{verbatim}
  (defthm row-echelon-p-id-fmat
    (implies (and (fmatp a n n) (posp n) (row-echelon-p a) (= (num-nonzero-rows a) n))
             (equal a (id-fmat n)))
\end{verbatim}
\end{small}
Now let

\begin{small}
\begin{verbatim}
  p = (inverse-mat a) = (row-reduce-mat a) = (row-ops-mat (row-reduce-ops a) n),

  q = (row-ops-mat (invert-row-ops (row-reduce-ops a)) n),
\end{verbatim}
\end{small}
and

\begin{small}
\begin{verbatim}
  r = (fmat* p a) = (row-reduce a).
\end{verbatim}
\end{small}
By {\tt invert-row-ops-mat}, {\tt (fmat* p q) = (fmat* q p) = (id-fmat n)}.  Suppose {\tt (row-rank r)} = {\tt n}.  By {\tt row-echelon-p-id-fmat},
{\tt (fmat* p a)} = {\tt r} = {\tt (id-fmat n)}, and by {\tt inverse-unique}, {\tt a = q}.  Thus, {\tt (invertiblep a n)} is a sufficient condition for
the existence of an inverse:

\begin{small}
\begin{verbatim}
  (defthmd invertiblep-sufficient
    (implies (and (fmatp a n n) (posp n) (invertiblep a n))
           (let ((p (inverse-mat a)))
               (and (fmatp p n n)
                    (equal (fmat* a p) (id-fmat n))
                    (equal (fmat* p a) (id-fmat n))))))
\end{verbatim}
\end{small}
To prove the necessity of {\tt (invertiblep a n)}, suppose {\tt (fmatp b n n)} and {\tt (fmat* a b) = (id\-fmat n)}.
Then

\begin{small}
\begin{verbatim}
  (fmat* r (fmat* b q)) = (fmat* (fmt* p a) (fmat* b q))
                        = (fmat* p (fmat* (fmat* a b) q))
                        = (fmat* p q)
                        = (id-fmat n).
\end{verbatim}
\end{small}
If {\tt (invertiblep a n) = NIL}, then the last row of {\tt r} is zero, and the same must be true of {\tt (id-fmat n)}, a contradiction.

\begin{small}
\begin{verbatim}
  (defthmd invertiblep-necessary
    (implies (and (fmatp a n n) (fmatp b n n) (posp n) (= (fmat* a b) (id-fmat n)))
             (invertiblep a n)))
\end{verbatim}
\end{small}

We note several consequences of the preceding results.  First, an invertible matrix is the inverse of its inverse:

\begin{small}
\begin{verbatim}
  (defthmd inverse-inverse-mat
    (implies (and (fmatp a n n) (posp n) (invertiblep a n))
             (and (invertiblep (inverse-mat a) n)
                  (equal (inverse-mat (inverse-mat a)) a))))
\end{verbatim}
\end{small}
Cancellation laws hold for invertible matrices, e.g.,

\begin{small}
\begin{verbatim}
  (defthmd invertiblep-cancel
    (implies (and (fmatp a m n) (fmatp b m n) (fmatp p m m) (posp m) (posp n)
                   (invertiblep p m))
             (iff (equal (fmat* p a) (fmat* p b))
                  (equal a b))))
\end{verbatim}
\end{small}
A matrix product is invertible iff each factor is invertible:

\begin{small}
\begin{verbatim}
  (defthmd invertiblep-factor
    (implies (and (fmatp a n n) (fmatp b n n) (posp n) (invertiblep (fmat* a b) n))
             (and (invertiblep a n) (invertiblep b n))))

  (defthmd inverse-fmat*
    (implies (and (fmatp a n n) (fmatp b n n) (posp n)
                  (invertiblep a n) (invertiblep b n))
             (and (invertiblep (fmat* a b) n)
                  (equal (inverse-mat (fmat* a b))
                         (fmat* (inverse-mat b) (inverse-mat a))))))
\end{verbatim}
\end{small}
Finally, we shall show that a is invertible iff its determinant is nonzero.  First note that if {\tt a} has inverse {\tt b}
and {\tt (fdet a)} = 0, then by {\tt fdet-multiplicative},

\begin{small}
\begin{verbatim}
  (fdet (id-fmat n) n) = (fdet (fmat* a b)) = (f* (fdet a) (fdet b)) = 0,
\end{verbatim}
\end{small}
a contradiction.  Thus,

\begin{small}
\begin{verbatim}
  (defthmd invertiblep-fdet-not-zero
    (implies (and (fmatp a n n) (posp n) (invertiblep a n))
             (not (equal (fdet a n) (f0)))))
\end{verbatim}
\end{small}
On the other hand, assume {\tt (fdet a n)} $\neq$ 0.  By {\tt fmat*-adjoint-fmat},

\begin{small}
\begin{verbatim}
  (fmat* a (adjoint-fmat a n)) = (fmat-scalar-mul (fdet a n) (id-fmat n)),
\end{verbatim}
\end{small}
which implies

\begin{small}
\begin{verbatim}
  (fmat* a (fmat-scalar-mul (f/ (fdet a n)) (adjoint-fmat a n)))
     = (fmat-scalar-mul (f/ (fdet a n)) (fmat* a (adjoint-fmat a n)))
     = (fmat-scalar-mul (f/ (fdet a n)) (fmat-scalar-mul (fdet a n) (id-fmat n)))
     = (id-fmat n),
\end{verbatim}
\end{small}
and by {\tt invertiblep-necessary}, {\tt a} is invertible.  This also establishes an alternative method for computing
the inverse:

\begin{small}
\begin{verbatim}
  (defthmd fdet-not-invertiblep-zero
    (implies (and (fmatp a n n) (natp n) (> n 1) (not (equal (fdet a n) (f0))))
             (and (invertiblep a n)
                  (equal (inverse-mat a)
                         (fmat-scalar-mul (f/ (fdet a n)) (adjoint-fmat a n))))))
\end{verbatim}
\end{small}

\section{Simultaneous Systems of Linear equations}\label{equations}

Let {\tt a} be an {\tt m}$\times${\tt n} matrix with {\tt (entry i j a)} = $a_{{\tt i},{\tt j}}$ for 0 $\leq$ {\tt i} $<$ {\tt m} and 0 $\leq$ {\tt j} $<$ {\tt n},
and let {\tt b} = ${\tt {}\tt (}b_0 \ldots b_{{\tt m}-1}{\tt ){\tt }}$ be an flist of length {\tt m}.
We seek an flist {\tt x} = ${\tt {}\tt (}x_0 \ldots x_{{\tt n}-1}{\tt ){\tt }}$ of length {\tt n} such that for 0 $\leq$ {\tt i} $<$ {\tt m},
\[
a_{{\tt i},0}x_0 + \ldots + a_{{\tt i},{\tt n}-1}x_{{\tt n}-1} = b_{\tt i}.
\]
We shall refer to {\tt a} as the {\it coefficient matrix} of this system of {\tt m} linear equations in {\tt n} unknowns.  To express the system as
a matrix equation, we define the {\it column matrix} corresponding to a given flist:

\begin{small}
\begin{verbatim}
  (defund col-mat (x) (transpose-mat (list x)))
\end{verbatim}
\end{small}
The above equations are naturally expressed by the matrix equation in the following definition:

\begin{small}
\begin{verbatim}
  (defund solutionp (x a b) (equal (fmat* a (col-mat x)) (col-mat b)))
\end{verbatim}
\end{small}
Let {\tt bc} = {\tt (col-mat b)}, {\tt xc} = {\tt (col-mat x)}, {\tt p} = {\tt (row-reduce-mat a)}, {\tt ar} = {\tt (fmat* p a)}, and {\tt br} = {\tt (fmat* p bc)}.
Left-multiplying the above equation by {\tt p} yields the equivalent equation

\begin{small}
\begin{equation}\label{eqn1}
  \mbox{\tt (fmat* ar xc) = br}.
\end{equation}
\end{small}
Thus, we have

\begin{small}
\begin{verbatim}
  (defthmd reduce-linear-equations
    (implies (and (fmatp a m n) (posp m) (posp n) (flistnp b m) (flistnp x n))
             (let* ((bc (col-mat b)) (xc (col-mat x))
                    (p (row-reduce-mat a)) (ar (fmat* p a)) (br (fmat* p bc)))
               (iff (solutionp x a b)
                    (equal (fmat* ar xc) br)))))
\end{verbatim}
\end{small}
Our objective, therefore, is to compute an {\tt n}$\times${\tt 1} column matrix {\tt xc} that solves Equation~(\ref{eqn1}), in which {\tt ar} is an {\tt m}$\times${\tt n}
reduced row-echelon matrix and {\tt br} is an {\tt m}$\times${\tt 1} column matrix.

Let {\tt q = (num-nonzero-rows ar) = (row-rank a)}.  We shall show that the existence of a solution to this equation is determined by whether the last {\tt m} $-$ {\tt q}
entries of {\tt br} are all 0.  This is true iff the following search returns {\tt NIL}:

\begin{small}
\begin{verbatim}
  (defun find-nonzero (br q m)
    (if (and (natp q) (natp m) (< q m))
        (if (= (entry (1- m) 0 br) (f0))
            (find-nonzero br q (1- m))
          (1- m))
      ()))
\end{verbatim}
\end{small}
Thus, we define

\begin{small}
\begin{verbatim}
  (defun solvablep (a b)
    (null (find-nonzero (fmat* (row-reduce-mat a) (col-mat b))
                        (row-rank a)
                        (len a))))
\end{verbatim}
\end{small}
Suppose first that {\tt (find-nonzero br q m)} = {\tt k} $\neq$ {\tt NIL}, so that {\tt (solvablep a b)} = {\tt NIL}.  Then 
{\tt (row k ar)} = {\tt (flistn0 n)} and {\tt (entry k 0 br)} $\neq$ 0.  It follows that {\tt (entry k 0 (fmat* ar xc))} $\neq$
{\tt (nth k 0 br)}, and hence {\tt (fmat* ar xc)} $\neq$ {\tt br}.  Combining this with {\tt reduce-linear-equations}, we
conclude that the system of equations has no solution:

\begin{small}
\begin{verbatim}
  (defthmd linear-equations-unsolvable-case
    (implies (and (fmatp a m n) (posp m) (posp n) (flistnp b m) (flistnp x n)
                  (not (solvablep a b)))
             (not (solutionp x a b))))
\end{verbatim}
\end{small}

Thus, we may assume {\tt (solvablep a b)} = {\tt T}.  As a first step toward the solution, consider the matrices
{\tt aq} and {\tt bq} consisting of the first {\tt q} rows of {\tt ar} and {\tt br}, respectively, computed by the
following:

\begin{small}
\begin{verbatim}
  (defun first-rows (q a)
    (if (zp q) ()
      (cons (car a) (first-rows (1- q) (cdr a)))))
\end{verbatim}
\end{small}
It is easily shown that {\tt aq} is a reduced row-echelon {\tt q}$\times${\tt n} matrix of row rank {\tt q} and that {\tt (fmat* ar xc)} = {\tt br}
iff {\tt (fmat* aq xc)} = {\tt bq}.  Our objective, therefore, is to solve the equation {\tt (fmat* aq xc) = bq}.

\subsection{Uniquely Solvable Case}

By {\tt row-rank<=n}, {\tt q} $\leq$ {\tt n}.  We first consider the case {\tt q} = {\tt n}.  By {\tt row-echelon-p-id-fmat},
{\tt aq} = {\tt (id-fmat n)} and {\tt (fmat* aq xc)} = {\tt bq} iff {\tt xc} = {\tt bq}.  Combining this observation
with {\tt first-rows\-linear-equations} and {\tt reduce-linear-equations}, we conclude that there exists a unique
solution in this case:;

\begin{small}
\begin{verbatim}
  (defthmd linear-equations-unique-solution-case
    (let* ((br (fmat* (row-reduce-mat a) (col-mat b)))
           (bq (first-rows n br)))
      (implies (and (fmatp a m n) (posp m) (posp n) (flistnp b m) (flistnp x n)
                    (solvablep a b) (= (row-rank a) n))
               (iff (solutionp x a b)
                    (equal x (col 0 bq))))))
\end{verbatim}
\end{small}

Our results on cofactor expansion lead to an alternative method of solving a system of {\tt n} linear 
equations in {\tt n} unknowns in the case of a unique solution, known as Cramer's rule.  Suppose {\tt m} = {\tt n}
= {\tt q}, so that {\tt a} is an invertible {\tt n}$\times${\tt n} matrix.  Our objective is to compute, as a function of {\tt a} and {\tt b},
for each {\tt i} $<$ {\tt n}, the {\tt i}th component {\tt (nth i x)} of the unique {\tt x} such that

\begin{equation}\label{eqn2}
    \mbox{\tt (fmat* a xc) = bc.}
\end{equation}
We refer to the analogs of the results of \cite[Sec. 5]{linear1} that appear in {\tt fdet.lisp}.  In particular, we shall substitute 
{\tt a'} = {\tt (replace-row (transpose-mat a) i b)} for {\tt a} in {\tt fdot-cofactor-fmat-row-fdet}.  Clearly, {\tt (row i a')} 
= {\tt b}.  By {\tt cofactor-fmat-transpose},

\begin{small}
\begin{verbatim}
  (cofactor-fmat-row i a' n) = (cofactor-fmat-row i (transpose-mat a) n)
                             = (row i (cofactor-fmat (transpose-mat a) n))
                             = (row i (adjoint-fmat a n)),
\end{verbatim}
\end{small}
and by {\tt fdet-transpose},

\begin{small}
\begin{verbatim}
  (fdet a' n) = (fdet (transpose-fmat (replace-col a i b)) n)
              = (fdet (replace-col a i b) n).
\end{verbatim}
\end{small}
Thus, the substitution yields the following:

\begin{small}
\begin{verbatim}
  (fdot b (row i (adjoint-fmat a n))) = (fdet (replace-col a i b) n)).
\end{verbatim}
\end{small}
Multiplying Equation~(\ref{eqn2}) by {\tt (adjoint-fmat a n)} yields

\begin{small}
\begin{verbatim}
  (fmat* (adjoint-fmat a n) (fmat* a xc)) = (fmat* (adjoint-fmat a n) bc).
\end{verbatim}
\end{small}
But

\begin{small}
\begin{verbatim}
  (fmat* (adjoint-fmat a n) (fmat* a xc))
    = (fmat* (fmat* (adjoint-fmat a n) a) xc)
    = (fmat* (flist-scalar-mul (fdet a n) (id-fmat n)) xc)
    = (flist-scalar-mul (fdet a n) (fmat* (id-fmat n) xc))         
    = (flist-scalar-mul (fdet a n) xc),
\end{verbatim}
\end{small}
and hence

\begin{small}
\begin{verbatim}
  (flist-scalar-mul (fdet a n) xc) = (fmat* (adjoint-fmat a n) bc).
\end{verbatim}
\end{small}
Equating the entries of these matrices in row {\tt i} and column 0, we have

\begin{small}
\begin{verbatim}
  (f* (fdet a n) (nth i x)) = (fdot b (row i (adjoint-fmat a n)))
                            = (fdet (replace-col a i b) n),
\end{verbatim}
\end{small}
which yields Cramer's rule:

\begin{small}
\begin{verbatim}
  (defthmd cramer
    (implies (and (fmatp a n n) (natp n) (> n 1) (invertiblep a n)
                  (flistnp b n) (flistnp x n) (solutionp x a b)
                  (natp i) (< i n))
             (equal (nth i x)
                    (f* (f/ (fdet a n))
                        (fdet (replace-col a i b) n)))))
\end{verbatim}
\end{small}

\subsection{General Solvable Case}

In the remainder of this section, we treat the general case {\tt (solvablep a b)} = {\tt T} with arbitrary
{\tt q} = {\tt (row-rank a)} $\leq$ {\tt n}.  The desired equation {\tt (fmat* aq xc)} = {\tt bq} holds iff for
0 $\leq$ {\tt i} $<$ {\tt q},

\begin{small}
\begin{verbatim}
                         (nth i (fmat* aq xc)) = (nth i bq)
\end{verbatim}
\end{small}
or equivalently,

\begin{small}
\begin{equation}\label{eqn3}
\mbox{\tt (fdot (row i aq) x) = (car (nth i bq)).}
\end{equation}
\end{small}
We shall split the dot product {\tt (fdot (nth i aq) x)} into two sums, corresponding to the list {\tt (lead\-inds aq)} of leading indices
and the list of remaining indices, which we call the {\it free indices}:

\begin{small}
\begin{verbatim}
  (defund free-inds (a n) (set-difference-equal (ninit n) (lead-inds a)))
\end{verbatim}
\end{small}
In general, given a sublist {\tt inds} of {\tt (ninit n)} and two flists {\tt r} and {\tt x} of length {\tt n}, the
following function extracts and sums the terms of the dot product of {\tt r} and {\tt x} that correspond to the indices
{\tt inds}:

\begin{small}
\begin{verbatim}
  (defun fdot-select (inds r x)
    (if (consp inds)
        (f+ (f* (nth (car inds) r) (nth (car inds) x))
            (fdot-select (cdr inds) r x))
      (f0)))
\end{verbatim}
\end{small}
In particular, {\tt (fdot (row i aq) x)} may be expressed as the following sum:

\begin{small}
\begin{verbatim}
    (f+ (fdot-select (lead-inds aq) (row i aq) x)
        (fdot-select (free-inds aq n) (row i aq) x))))).
\end{verbatim}
\end{small}
Now since {\tt (row i aq)} has a 1 at index {\tt (nth i (lead-inds aq))} and a 0 at all other lead indices, the first
of these two sums reduces to the single term {\tt (nth (nth i (lead-inds aq)) x)}, and hence Equation~(\ref{eqn3}) may be
expressed as

\begin{small}
\begin{verbatim}
  (nth (nth i (lead-inds aq)) x)
    = (f+ (car (nth i bq))
          (f- (fdot-select (free-inds aq n) (row i aq) x))).
\end{verbatim}
\end{small}
Thus, {\tt x} is a solution of our system of equations iff this condition holds for all {\tt i} $<$ {\tt q}.  This
is checked recursively by the following function:

\begin{small}
\begin{verbatim}
  (defun solution-test-aux (x aq bq lead-inds free-inds k)
    (if (zp k) t
      (and (equal (nth (nth (1- k) lead-inds) x)
                  (f+ (car (nth (1- k) bq))
                      (f- (fdot-select free-inds (nth (1- k) aq) x))))
           (solution-test-aux x aq bq lead-inds free-inds (1- k)))))

  (defund solution-test (x a b n)
    (let* ((ar (row-reduce a))
           (br (fmat* (row-reduce-mat a) (col-mat b)))
           (q (num-nonzero-rows ar))
           (aq (first-rows q ar))
           (bq (first-rows q br))
           (lead-inds (lead-inds aq))
           (free-inds (free-inds aq n)))
      (solution-test-aux x aq bq lead-inds free-inds q)))
  
\end{verbatim}
\end{small}
This provides a test to be applied to a candidate solution:

\begin{small}
\begin{verbatim}
  (defthmd linear-equations-solvable-case
    (implies (and (fmatp a m n) (posp m) (posp n) (flistnp b m) (flistnp x n)
                  (solvablep a b))
             (iff (solutionp x a b)
                  (solution-test x a b n))))
\end{verbatim}
\end{small}
If {\tt q} = {\tt (len (lead-inds aq))} = {\tt n}, then {\tt (free-inds aq n)} = {\tt NIL},  the equation

\begin{small}
\begin{verbatim}
  (nth (nth i l) x) = (f+ (car (nth i bq)) (f- (fdot-select f (nth i aq) x)))
\end{verbatim}
\end{small}
reduces to

\begin{small}
\begin{verbatim}
  (nth i x) = (car (nth i bq),
\end{verbatim}
\end{small}
{\tt (solution-test-aux x aq bq l f q)} reduces to {\tt x = (col 0 bq)}, and the last theorem reduces to the earlier result
{\tt linear-equations-unique-solution-case}.

Otherwise, {\tt (free-inds aq n)} $\neq$ {\tt NIL} and the components of {\tt x} corresponding to
the indices in {\tt (lead-inds aq)} are determined by the components corresponding to {\tt (free-inds aq n)}, which are unconstrained.
Thus, there is a single solution corresponding to every assignment of values to the latter set of components, and hence infinitely
many solutions.  We shall revisit this result in Part III, where we show that in the homogeneous case, {\tt b} = {\tt (flistn0 n)}, the
solutions form a vector space of dimension {\tt n} $-$ {\tt q}.  A basis for this solution space will be provided by a formula
derived from the function {\tt solution-test}.
.

\section{Future Work}

This formalization of linear algebra is a work in progress.  In Part~I, we developed the algebra of matrices over a commutative
ring with unity and the theory of determinants.  In this sequel, we have restricted our attention to matrices over a field in
order to address the process of row reduction and its applications.  To allow our results to be applied to an arbitrary ring
or field, we have characterized each by an encapsulated set of constrained functions.

There is progress to report on a planned Part~III, which begins with another encapsulation that formalizes the notion of an abstract
finite-dimensional vector space over the field {\tt F}.  The constrained functions of this encapsulation naturally include a
predicate that recognizes vectors in the space, the operations of vector addition and scalar multiplication, and the constant 0
vector.  Two additional functions embody the requirement of finite dimensionality: (1)~a constant list of vectors of unspecified
length that serves as a canonical basis, and (2)~a function that returns the coordinates of  a given vector with respect to this
basis.  Thus, whenever we define a concrete vector space, we are obligated to identify a basis for it.  This establishes a tight
connection between vector spaces and matrices: a list of vectors may be identified by the matrix of coordinates of its members.
As an unexpected application of row reduction, this connection provides an algorithmic definition of the basic notion of linear
independence without use of quantifiers: a list of vectors is linearly independent iff the row rank of its coordinate matrix is
the length of the list.

The reader may have noticed that the definition of the basic notion of row equivalence is omitted from our treatment of row
reduction.  Recall that two matrices are said to be row equivalent if one may be derived from the other by a sequence of
elementary row operations.  This definition could be formalized in ACL2 using the support for existential quantification provided
by {\tt defun-sk}, and the properties of an equivalence relation could be derived from the results of Section~\ref{reduction}.
We could also prove the important theorem that distinct reduced row-echelon matrices cannot be row equivalent.  However, since
its most expedient proof is based on vector spaces (in particular, the row space of a matrix), this result is postponed to Part
III.  Note that it provides an alternative definition of row equivalence that avoids quantification: two matrices {\tt a} and
{\tt b} are row equivalent iff {\tt (row-reduce a)} = {\tt (row-reduce b)}.  Since we prefer this algorithmic formulation, the
entire topic is deferred to Part III.

Other topics to be addressed in the sequel include linear transformations and diagonalization.  As discussed in
Part~I, a factor in our decision to develop the algebra of matrices over an arbitrary commutative ring rather than a field
is that this allows us to define the characteristic polynomial of a square matrix over the field {\tt F} as the determinant of a
certain matrix over the polynomial ring ${\tt F}[t]$.  A related objective is the proof of the Cayley-Hamilton Theorem (every
square matrix over a commutative ring is a root of its own characteristic polynomial), which has wide-ranging applications in
other areas of mathematics.

This project is part of a broader effort in the formalization of algebra, which began with group theory \cite{groups1, groups2,
groups3} and will continue beyond linear algebra.  Our next targeted area of investigation will be Galois theory, which we hope
eventually to apply to the study of algebraic number fields.  These intended applications guided our choices of formalization
schemes for the basic algebraic structures.  Since we are interested in infinite rings and fields, the encapsulation approach
seems to be the only viable representation scheme for these structures provided by the ACL2 logic.  The disadvantages of not
being able to refer to such a structure as an ACL2 object are obvious.  On the other hand, since the groups of primary interest
are finite, our investigation of group theory is limited to the finite case.  Under this restriction, a group is conveniently
represented as an object characterized by a predicate defined as an ACL2 function.  There are, however, infinite groups of
interest, which are not accommodated by our theory.  For example, the general linear group of invertible matrices over a field,
which is in general an infinite structure, would otherwise have provided an interesting example and another connection between
group theory and linear algebra.

Although we have no immediate plans to apply this theory beyond the realm of pure mathematics, its potential utility in the formal
verification of hardware and software applications is limitless.  Linear algebra is central to the rapidly advancing fields of
machine learning and neural networks \cite{halim}, providing essential tools for data reprersentation and manipulation.  Along
with finite group theory, it is also important to a variety of cryptographic algorithms \cite{qian}.  It is our hope that ACL2
users who are interested in pursuing such applications may find our results useful.

\nocite{*}
\bibstyle{eptcs}
\bibliographystyle{eptcs}
\bibliography{linear2}
\end{document}